\documentclass{article}
\usepackage{geometry}
\usepackage{tikz}
\usepackage{subfigure}
\usetikzlibrary{calc}
\usetikzlibrary{arrows}
 \usepackage{relsize}

\newcommand{\added}[1]{#1}
\newcommand{\addedtwo}[1]{#1}

\usepackage{amsmath}
\usepackage{amssymb}
\usepackage{amsfonts}
\usepackage{esint}
\usepackage{mathrsfs}
\usepackage{amsthm}
\usepackage{amscd}
\usepackage{longtable}
\usepackage{listings}
\usepackage{booktabs,longtable}
\usepackage{graphicx}
\usepackage{multirow}

\newcommand{\ppulp}[1]{\mathsf{#1}}

\newcommand{\ppulpm}[1]{\pmb{\ppulp{#1}}}

\let \oldcdot \cdot
\renewcommand{\oldcdot}[1][0pt]{%
  \mathrel{\raisebox{#1}{{\Large $\cdot$}}}%
}
\newcommand*{\xdash}[1][3em]{\rule[0.5ex]{#1}{0.55pt}}

\newcommand{\vect}[1]{\mathbf{#1}}

\newcommand{\matr}[1]{\mathbf{#1}}
\newcommand{\abs}[1]{\left \lvert #1 \right\rvert }

\renewcommand{\Re}[1]{\mathbb{R}\mathrm{e}\left \{#1\right\} }
\renewcommand{\Im}[1]{\mathbb{I}\mathrm{m}\left \{#1\right\} }

\newcommand{\pref}[1]{(\ref{#1})}
\newcommand{\junk}[1] {}

\def\XXint#1#2#3{{\setbox0=\hbox{$#1{#2#3}{\int}$}
\vcenter{\hbox{$#2#3$}}\kern-.5\wd0}}

\newcommand*\widebar[1]{%
  \hbox{%
    \vbox{%
      \hrule height 0.5pt 
      \kern0.3ex
      \hbox{%
        \kern-0.05em
        \ensuremath{#1}%
        \kern-0.05em
      }%
    }%
  }%
}

\begin{document}
%

\title{MoM-SO: a Complete Method for Computing the Impedance of Cable Systems Including Skin, Proximity, and Ground Return Effects}

\author{Utkarsh~R.~Patel
        and~Piero~Triverio
\thanks{Manuscript received ...; revised ...}%
\thanks{This work was supported in part by the KPN project "Electromagnetic transients in future power systems" (ref. 207160/E20) financed by the Norwegian Research Council (RENERGI programme) and by a consortium of industry partners led by SINTEF Energy Research: DONG Energy, EdF, EirGrid, Hafslund Nett, National Grid, Nexans Norway, RTE, Siemens Wind Power, Statnett, Statkraft, and Vestas Wind Systems.}
\thanks{U.~R.~Patel and P.~Triverio are with the Edward S. Rogers Sr. Department of Electrical and Computer Engineering, University of Toronto, Toronto, M5S 3G4 Canada (email: utkarsh.patel@mail.utoronto.ca, piero.triverio@utoronto.ca).}\\[2em]
This paper is published in the IEEE Transaction on Power Delivery \\ Oct. 2015, vol. 30, no. 5, pp. 2110-2118.
DOI: 10.1109/TPWRD.2014.2378594
}

\date{}
%



\maketitle

\begin{abstract}
The availability of accurate and broadband models for underground and submarine cable systems is of paramount importance for the correct prediction of electromagnetic transients in power grids. Recently, we proposed the MoM-SO method for extracting the series impedance of power cables while accounting for skin and proximity effect in the conductors. In this paper, we extend the method to include ground return effects and to handle cables placed inside a tunnel. Numerical tests show that the proposed method is more accurate than widely-used analytic formulas, and is much faster than existing proximity-aware approaches like finite elements. For a three-phase cable system in a tunnel, the proposed method requires only 0.3~seconds of CPU time per frequency point, against the 8.3~minutes taken by finite elements, for a speed up beyond 1000~X. 
\end{abstract}

\section{Introduction} 

Electromagnetic transients are a growing concern in the design and operation of power systems. Their prediction using Electro-Magnetic Transient (EMT) programs like~\cite{Cho96,Ame13} requires broadband models for each component of the power system, including underground and submarine cables~\cite{Mor99, Mar82, Nod96}. In order to create a cable model for transient analysis, we require the per-unit-length (p.u.l.) series impedance of the cable over the frequency range of interest, which typically extends from a few Hz to the MHz range. 
The broadband p.u.l. parameters of the cable must account for frequency-dependent phenomena that take place inside the cable, namely skin and proximity effects. Moreover, for buried cables, they must also take into account the return current that may flow in the surrounding soil.

Existing EMT tools use analytic formulas~\cite{Ame80, Wed73} to compute the series impedance of cables. Such formulas include skin effect but neglect proximity effects which are significant in closely-packed cables, where conductors' proximity leads to a non-uniform current distribution in the conductors. For buried cables, the contribution to the impedance due to ground return is added through Pollaczek's formula~\cite{Pol26}. Since Pollaczek's formula involves an infinite integral, a series approximation due to Saad~\cite{Saad96} is typically preferred. \added{This approach, however, is not accurate at high frequency for certain cable configurations, as our numerical tests will show}. Additionally, Pollaczek's formula neglects proximity effects inside ground, and cannot account for the presence of a tunnel around the cable. The limitations of analytic formulas can be overcome using finite elements (FEM)~\cite{Wei82, B09, Cri89, Hab13} or conductor partitioning~\cite{Ame92, Com73, Dea87, Pag12}. These approaches correctly capture skin, proximity and ground effects~\cite{Yin89}. However, they can be very time consuming. Since ground is a poor conductor, at low frequency skin depth in earth can be as high as 5~km\footnote{at 1~Hz and for a soil conductivity of 0.01 S/m.}. Therefore, the FEM mesh must extend over a huge domain in order to correctly predict losses in ground. Moreover, as frequency grows and skin depth becomes very small, one is forced to remesh the geometry in order to correctly model current crowding near conductors' boundaries. These issues make a FEM analysis very time consuming and impractical for a power engineer that typically does not have a deep expertise in finite element methods. The development of a fast and easy-to-use method to accurately characterize power cables is the objective of this research.

In~\cite{TPWRD1, TPWRD2}, we proposed an efficient and proximity-aware method, dubbed MoM-SO, to compute the series impedance of cables with round conductors, both solid and hollow (tubular). In this technique, conductors are represented through an equivalent current placed on their surface. Using a surface admittance operator~\cite{DeZ05} and the Green's function of the surrounding medium, this representation allows for the computation of the cable impedance. This approach is faster than finite elements or conductor partitioning because it does not require a meshing of the whole cross section of the cable system, but only a discretization of the conductors' boundary. In this paper, we extend our previous work~\cite{TPWRD2} in two directions. Firstly, we fully include ground return effects, which were only taken into account in an approximate way in~\cite{TPWRD2}. Secondly, the proposed method can handle cables placed inside one or multiple holes or tunnels dug in ground. In order to account for the effect of the hole/tunnel on the cable impedance, we introduce a surface admittance representation for the cable-hole system, which is a novel result and makes the computation very efficient.

The paper is organized as follows. After formulating the problem in Sec.~\ref{sec:formulation}, we develop the surface admittance operator for the cable-hole system in Sec.~\ref{sec:SAoperator}. In Sec.~\ref{sec:GroundReturn}, the effect of ground conductivity is introduced using the Green's function of the air-ground medium, and in Sec.~\ref{sec:efie} the p.u.l. cable impedance is obtained. Finally, in Sec.~\ref{sec:Results} the proposed MoM-SO method is compared against a commercial FEM solver~\cite{COMSOL} and analytic formulas. Numerical tests demonstrate the excellent accuracy and computational efficiency of MoM-SO.

\section{Problem Formulation}
\label{sec:formulation}

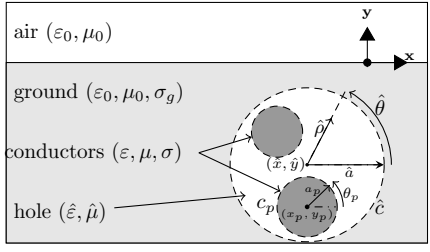
\begin{figure}
\centering
\begin{tikzpicture}[scale=0.8, every node/.style={scale=0.8}]
\draw [fill=white] (-1,4) rectangle (6,0) node (v1) {};
\draw  [fill=black!10](-1,3) rectangle (v1);

\node at (3,3.5) {};
\node at (3,2.8) {};

\draw  [ fill=white, densely dashed] (4,1.3) ellipse (1.28 and 1.28) node [left] {};
\draw  [fill=black!40, densely dashed] (3.5,1.85) ellipse (0.425 and 0.425) node {};
\node at (3.5,1.7) [above] {};

\draw  [fill = black!40,densely dashed] (4,0.6) ellipse (0.5 and 0.5) node [below] {} ;
\draw[fill=black](4,0.6) ellipse(0.01 and 0.01) node [above]{};
\node (v4) at (4,0.52)  [above] {};

\node at (4.5,2.2) [below] {};

\node (v2) at (5,2.8) {};
\node (v6) at (5,3.0) {};
\node (v5) at (4.8,3.0) {};
\node (v3) at (5,3.8) {\scriptsize ${\bf y}$};
\draw [-triangle 60] (v2) edge (v3);

\node (v4) at (5.8,3)  {};
\node(x1) at (5.7,3.1) {\scriptsize ${\bf x}$};
\draw [-triangle 60] (v5) edge (v4);

\draw  [ fill=black] (v6) ellipse (0.06 and 0.06);
\draw  [ fill=black] (4,1.3) ellipse (0.02 and 0.02) node [left] (v8) {};
\node at (4.1,1.3) [left] {\scriptsize $(\hat{x}, \hat{y})$};

\node (v7) at (5.4,1.3) { };
\node at (4.7, 1.4) [below] {\scriptsize $\hat{a}$};
\draw[-stealth]   (4,1.3) -> (v7);

\node  at (5.2,0.6) {$\hat{c}$};

\node (center) at (4,1.3) {};
\node (center2) at (3.95,1.2) {};
\node (p1) at (5.4,1.3) {};
\node (p2) at (4.70,2.65) {};
\node(p12) at (4.47,2.2){};
\node (p3) at (4.2,1.65) {};
\node (p4) at (5.0, 2.3)  {};
\node (p5) at (5.4, 2.9) {};
\draw  [densely dashed] (center) -- (p2) {};
\draw  [->] (center2) -- (p12) {};

\node at (p3) [above] {\small $\hat{\rho}$};
\node at (p4) [right] {$\hat{\theta}$};
\draw [dashdotted] (center) -- (p1) {};
\draw [->] (p1) arc (0:60:1.4);

\node (p6) at (4, 0.6)  {};
\node (p66) at (3.9, 0.5)  {};
\node (p7) at (4.6, 0.6){};
\node(p8) at (4.5,1.08){};
\node(p9) at (4.25,1.4){};
\node(p10) at (4.75,1.5){};

\node at (4.1,0.85){\tiny$a_p$};

\draw [dashdotted] (p6) -- (p7){};
\draw [->] (p7) arc (0:45:0.6);
\draw [->] (p66) -- (p8){};
\draw  [fill=black] (p6) ellipse (0.03 and 0.03);

\node (p14) at (3.6,0.6) [left] {$c_p$};

\node (p15) at (4.47,0.8) [right] {\scriptsize $\theta_p$};
\node at (4,0.68) [below] {\tiny$(x_p, y_p)$};

\node at (-1,3.5) [right] {air ($\varepsilon_0, \mu_0$)};

\node at (-1,2.5) [right] {ground ($\varepsilon_0, \mu_0, \sigma_g$)};
\node  (p17) at (2.05,1.5) [left] {conductors ($\varepsilon, \mu, \sigma$)};
\node (p18) at (3.2,1.8) {};
\node (p19) at (3.68,0.8) {};

\draw [->] (2.2,1.5)--(p18);
\draw [->] (2.2,1.5)--(p19);

\node (p20) at (-1,0.5) [right] {hole ($\hat{\varepsilon}, \hat{\mu}$)};
\node (p21) at (3.1,0.8) {};
\draw [->] (p20)--(p21);
\end{tikzpicture}
\caption{Cross-section of a simple cable with two conductors used to illustrate the proposed method. Notation for the conductivity, permittivity and permeability of each element is established. The coordinate system used in the paper is also presented.}
\label{fig:originalfig}
\end{figure}

Our goal is to compute the p.u.l. impedance, \added{as defined in \cite{Pau07}}, of a cable system made by round metallic conductors buried into one or multiple holes dug in a conductive soil.  
A simple configuration is depicted in Fig.~\ref{fig:originalfig}, and will be used to describe the MoM-SO technique. For the sake of clarity, we will develop the  theory behind MoM-SO considering only solid conductors and a single hole. However, as discussed in Sec.~\ref{sec:MultipleHole}, the proposed method can handle both solid conductors and hollow (tubular) conductors, placed in one or multiple holes excavated in ground. Hollow conductors are useful to model screens and armouring structures found, for example, in pipe-type cables.

We denote with $P$ the number of conductors present in the cable system.  As shown in Fig.~\ref{fig:originalfig}, the $p$-th conductor is centered at $(x_p, y_p)$ and has radius $a_p$.  Each conductor has electric permittivity $\varepsilon$, magnetic permeability $\mu$, and conductivity $\sigma$.  Although, for simplicity of notation, we assume here that these properties are the same for all conductors, the proposed method can handle different conductive materials with obvious modifications. Conductors are located inside a round hole, which is centered at $(\hat{x}, \hat{y})$ and whose radius is $\hat{a}$. The space inside the hole is lossless with permittivity $\hat{\varepsilon}$ and permeability $\hat{\mu}$. The background medium consists of air for $y>0$ and of a lossy soil of conductivity $\sigma_g$ for $y<0$. Both air and ground have permittivity $\varepsilon_0$ and permeability $\mu_0$. 

We are interested in computing the p.u.l. resistance $\ppulpm{R}(\omega)$ and inductance $\ppulpm{L}(\omega)$ matrices that relate the potential $V_p$ of each conductor to the current $I_p$ flowing in each conductor as
\begin{equation}
\frac{\partial \vect{V}}{\partial z} = - \left[\ppulpm{R}(\omega) + j \omega \ppulpm{L}(\omega) \right] \vect{I} \,,
\label{eq:Tel1}
\end{equation}
where vectors $\vect{V} = \begin{bmatrix} V_1 & V_2 & \hdots & V_P \end{bmatrix}^T$ and $\vect{I} = \begin{bmatrix} I_1 & I_2 & \hdots & I_P \end{bmatrix}^T$ store, respectively, the potential and current of each conductor.
\added{In our approach, the cable parameters are computed assuming that the electromagnetic field is longitudinally invariant along the cable, neglecting ``end effects''. These effects may be relevant for short cables~\cite{Ame05}. In order to account for them, a 3D formulation must be used, increasing dramatically the computational cost. For this reason, our method is based on transmission line theory, which is extensively used in  cable modeling~\cite{Wed73, Pol26, Yin89, Pag12}. For a discussion on end effects, we point the Reader to~\cite{Ame05}.}

\section{Surface Admittance Representation for the Cable-Hole System}
\label{sec:SAoperator}

\subsection{Surface Admittance Representation for the Conductors}
In order to compute the p.u.l. impedance of the cable, we adopt the surface admittance approach of~\cite{DeZ05}. Firstly, we represent each conductor with an equivalent current on its surface. Then, the same operation will be performed on the hole boundary, leading to a very compact and efficient representation for the hole-cable system.  We let
\begin{equation}
\vect{r}_p(\theta_p) = (x_p + a_p \cos \theta_p)\, {\vect{x}} + (y_p + a_p \sin \theta_p )\, {\vect{y}}
\end{equation}
be the position vector which traces the contour $c_p$ of conductor $p$, as shown in Fig.~\ref{fig:originalfig}.
We expand the longitudinal electric field on the contour $c_p$ in a truncated Fourier series 
\begin{equation}
E_z(\theta_p) = \sum_{n=-N_p}^{N_p} E_n^{(p)} \operatorname{e}^{jn\theta_p} \,,
\label{eq:Eexp1}
\end{equation}
where $N_p$ controls the number of basis functions used to represent the field on the boundary. Numerical tests show that a $N_p$ of 3 or 4 is typically sufficient to accurately represent the electrical field in the conductors of a power cable~\cite{spi2013}. The number of basis functions $N_p$ can be determined automatically as discussed in~\cite{spi2013}.

\begin{figure}
\centering
\begin{tikzpicture}[scale=0.8, every node/.style={scale=0.8}]
\draw [fill=white] (2,4) rectangle (6,0) node (v1) {};
\draw  [fill=black!10](2,3) rectangle (v1);

\node at (3,3.5) {$\mu_0$, $\varepsilon_0$};
\node at (3,2.8) {$\sigma_g$, $\mu_0$, $\varepsilon_0$};

\draw  [ fill=white] (4,1.3) ellipse (1.28 and 1.28);
\draw  [dashdotted] [fill=white] (3.5,1.85) ellipse (0.42 and 0.42) node {$\hat{\mu}$, $\hat{\varepsilon}$};
\foreach \a in {20, 80,...,320} {
      \draw ({3.5+0.42*cos(\a)},{1.85+0.42*sin(\a)}) circle [radius=.08];
      \draw [fill=black] ({3.5+0.42*cos(\a)},{1.85+0.42*sin(\a)}) circle [radius=.04];
}


\draw  [dashdotted] [fill = white] (4,0.6) ellipse (0.5 and 0.5) node  {} ;
\foreach \a in {20, 80,...,320} {
      \draw ({4+0.5*cos(\a)},{0.6+0.5*sin(\a)}) circle [radius=.08];
      \draw [fill=black] ({4+0.5*cos(\a)},{0.6+0.5*sin(\a)}) circle [radius=.04];
}

\node at (4.5,2.2) [below] {$\hat{\mu}$, $\hat{\varepsilon}$};

\node at (4.5,2.2) [below] {};
\node at (4.7,1.35) { $J_s^{(p)}(\theta_p)$};


\foreach \a in {20, 80,...,320} {
}

\node (v2) at (5,2.8) {};
\node (v6) at (5,3.0) {};
\node (v5) at (4.8,3.0) {};
\node (v3) at (5,3.8) {\scriptsize $ {\bf y}$};
\draw [-triangle 60] (v2) edge (v3);

\node (v4) at (5.8,3)  {};
\node(x1) at (5.7,3.1) {\scriptsize ${\bf x}$};
\draw [-triangle 60] (v5) edge (v4);

\draw  [ fill=black] (v6) ellipse (0.06 and 0.06);


\end{tikzpicture}
\begin{tikzpicture}[scale=0.8, every node/.style={scale=0.8}]
\draw [fill=white] (2,4) rectangle (6,0) node (v1) {};
\draw  [fill=black!10](2,3) rectangle (v1);

\node at (3,3.5) {$\mu_0$, $\varepsilon_0$};
\node at (3,2.8) {$\sigma_g$, $\mu_0$, $\varepsilon_0$};

\draw  [dashdotted, fill=black!10] (4,1.3) ellipse (1.28 and 1.28)  node [below] {$\sigma_g, \mu_0$, $\varepsilon_0$};
\foreach \a in {20, 80,...,320} {
      \draw ({4+1.28*cos(\a)},{1.3+1.28*sin(\a)}) circle [radius=.08];
      \draw [fill=black] ({4+1.28*cos(\a)},{1.3+1.28*sin(\a)}) circle [radius=.04];
}

\node at (4.5, 2.1) {$\widehat{J}_s(\hat{\theta})$};



\node (v2) at (5,2.8) {};
\node (v6) at (5,3.0) {};
\node (v5) at (4.8,3.0) {};
\node (v3) at (5,3.8) {\scriptsize ${\bf y}$};
\draw [-triangle 60] (v2) edge (v3);

\node (v4) at (5.8,3)  {};
\node (x1) at (5.7,3.1) {\scriptsize $ {\bf x}$};
\draw [-triangle 60] (v5) edge (v4);

\draw  [ fill=black] (v6) ellipse (0.06 and 0.06);

\node  at (5.2,0.6) {$\hat{c}$};

\end{tikzpicture}
\caption{Left panel: cross-section of the cable in Fig.~\ref{fig:originalfig} after all conductors have been replaced by the surrounding hole medium. Equivalent currents $J_s^{(p)}(\theta_p)$ are introduced on their contours. Right panel: cross-section of the cable after application of the equivalence theorem to the boundary of the hole. An equivalent current $\widehat{J}_s(\hat{\theta})$ is introduced on the hole boundary $\hat{c}$.}
\label{fig:replaceconfig}
\end{figure}
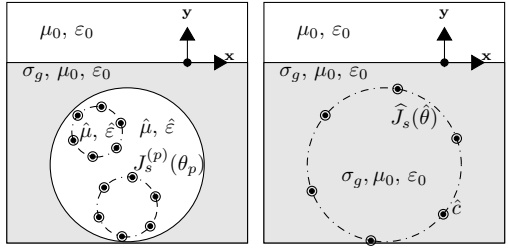

We now replace each conductor with the surrounding hole medium, introducing an equivalent current $J_s^{(p)}(\theta_p)$ on its boundary, as shown in the left panel of Fig.~\ref{fig:replaceconfig}. If $J_s^{(p)}(\theta_p)$ is chosen according to the equivalence theorem~\cite{Bal89}, this operation does not change the fields outside the conductors, allowing for the extraction of the p.u.l parameters.

The equivalent current on $c_p$ is also expanded in a truncated Fourier series
\begin{equation}
J_s^{(p)}(\theta_p) = \frac{1}{2\pi a_p} \sum_{n=-N_p}^{N_p} J_n^{(p)} \operatorname{e}^{jn\theta_p} \,.
\label{eq:Jexpan1}
\end{equation} Equivalence principle imposes the following relation~\cite{DeZ05} between the Fourier coefficients of electric field~\pref{eq:Eexp1} and surface current~\pref{eq:Jexpan1} 
\begin{equation}
J^{(p)}_{n} = E^{(p)}_{n} \frac{2\pi}{j\omega } \biggl [ \frac{k a_p {\cal J}_{|n|}'(ka_p)}{\mu {\cal J}_{|n|}(ka_p)} - \frac{\hat{k} a_p {\cal J}_{|n|}'(\hat{k} a_p)}{\hat{\mu} {\cal J}_{|n|}(\hat{k} a_p)}  \biggl ]\,,
\label{eq:JCoeff}
\end{equation}
where ${\cal J}_{|n|}(.)$ is the Bessel function of the first kind~\cite{Abr64} of order $|n|$, and ${\cal J}_{|n|}'(.)$ is its derivative. The quantities $k = \sqrt{ \omega \mu (\omega \varepsilon -j \sigma)}\, $  and $\hat{k} = \omega \sqrt{\hat{\mu} \hat{\varepsilon}}\,$ denote, respectively, the wavenumber inside the conductors and inside the hole.

If we collect the Fourier coefficients $E_n^{(p)}$ and  $J_n^{(p)}$ of all conductors into two column vectors
\begin{align}
\vect{E} &= \begin{bmatrix} E_{-N_1}^{(1)} & E_{-N_1+1}^{(1)} & \hdots & E_{N_1}^{(1)} & E_{-N_2}^{(2)} & \hdots \end{bmatrix}^T \,,\\
\vect{J} &= \begin{bmatrix} J_{-N_1}^{(1)} & J_{-N_1+1}^{(1)} & \hdots & J_{N_1}^{(1)} & J_{-N_2}^{(2)} & \hdots \end{bmatrix}^T \,,
\end{align}
we can compactly write~\pref{eq:JCoeff} as
\begin{equation}
\vect{J} = \matr{Y}_s \vect{E} \,,
\label{eq:surf1}
\end{equation}
where matrix $\matr{Y}_s$ can be interpreted as a surface admittance operator which relates the equivalent current~\pref{eq:Jexpan1} on the conductors to the corresponding electrical field~\pref{eq:Eexp1}.
Details on the surface admittance matrix $\matr{Y}_s$ can be found in \cite{TPWRD1}. At this point, we have considerably simplified the geometry of the problem and obtained the configuration shown in  the left panel of Figure~\ref{fig:replaceconfig}.

\subsection{Surface Admittance  Representation for the Cable-Hole System}
\label{sec:SAcablehole}

We next show that it is possible to further simplify the problem at hand by representing the entire cable-hole system with a unique equivalent current density $\widehat{J}_s(\hat{\theta})$ placed on the hole's boundary, as shown in Fig.~\ref{fig:replaceconfig} (right panel). The boundary of the hole is denoted by $\hat{c}$ and can be described by the position vector $\widehat{\vect{r}}(\hat{a}, \hat{\theta})$ where
\begin{equation}
\widehat{\vect{r}}(\hat{\rho}, \hat{\theta}) = \left( \hat{x} + \hat{\rho} \cos \hat{\theta}  \right) \vect{x} + \left( \hat{y} + \hat{\rho} \sin \hat{\theta} \right) \vect{y} \,,
\end{equation}
for $\hat{\rho} \in [0,\hat{a}]$, and $\hat{\theta} \in [0, 2\pi]$.

Similarly to our approach for round conductors, we represent the magnetic vector potential on the boundary of the hole with a truncated Fourier expansion
\begin{equation}
\widehat{A}_z(\hat{\theta}) = \sum_{n=-\widehat{N}}^{\widehat{N}} \widehat{A}_n \operatorname{e}^{jn\hat{\theta}} \,.
\label{eq:BC1}
\end{equation}
The coefficients of this expansion are cast into vector $\vect{\widehat{A}} = \begin{bmatrix} \widehat{A}_{-\widehat{N}} & \hdots & \widehat{A}_{\widehat{N}} \end{bmatrix}^T$. 
We replace the hole medium and all the equivalent currents inside it by the surrounding ground medium, as shown in Fig.~\ref{fig:replaceconfig} (right panel). 
In order to keep the fields outside of the hole \emph{unchanged}, we introduce an equivalent current
\begin{equation}
\widehat{J}_s(\hat{\theta}) = \frac{1}{2\pi \hat{a}} \sum_{n=-\widehat{N}}^{\widehat{N}} \widehat{J}_n \operatorname{e}^{jn\hat{\theta}} \,,
\label{eq:JHat1}
\end{equation}
on the hole boundary $\hat{c}$. The coefficients of $\widehat{J}_s(\hat{\theta})$ are stored in vector $\widehat{\vect{J}} = \begin{bmatrix} \widehat{J}_{-\widehat{N}} & \hdots & \widehat{J}_{\widehat{N}} \end{bmatrix}^T$. From the equivalence principle~\cite{Bal89}, it follows that the equivalent current must read
\begin{equation}
\widehat{J}_s (\hat{\theta}) = \left[ \frac{1}{\mu_0} \frac{\partial \widebar{\cal A}_z(\hat{\rho}, \hat{\theta)}}{\partial \hat{\rho}} - \frac{1}{\hat{\mu}} \frac{\partial \widehat{\cal A}_z(\hat{\rho},\hat{\theta})}{\partial \hat{\rho}}  \right]_{\hat{\rho} = \hat{a}} \,,
\label{eq:Jhateq1}
\end{equation}
where $\widehat{\cal A}_z(\hat{\rho}, \hat{\theta})$ is the longitudinal magnetic potential inside the hole in the configuration shown in the left panel of Fig.~\ref{fig:replaceconfig}. 
Instead, $\widebar{\cal A}_z(\hat{\rho}, \hat{\theta})$ is the magnetic potential inside the hole in the configuration shown in the right panel of Fig.~\ref{fig:replaceconfig}, i.e. after application of the equivalence theorem.

In order to evaluate~\pref{eq:Jhateq1}, we must determine the magnetic potential inside the hole. We first find the magnetic potential $\widehat{\cal A}_z(\hat{\rho}, \hat{\theta})$, which must satisfy the non-homogeneous Helmholtz equation~\cite{Bal89} 
\begin{align}
&\nabla^2 \widehat{\cal A}_z + \hat{k}^2 \widehat{\cal A}_z = -\hat{\mu} \sum_{q=1}^{P} J_s^{(p)}(\theta_p) \, \label{eq:PDE2} 
\end{align}
subject to the Dirichlet boundary condition \pref{eq:BC1} on $\hat{c}$.
The forcing term in~\pref{eq:PDE2} is the sum of all equivalent currents inside the hole. 
The solution of~\pref{eq:PDE2} can be written as the sum of the general solution $\widehat{\cal A}_z'(\hat{\rho}, \hat{\theta})$ and the particular solution  $\widehat{A}_z''(\hat{\rho}, \hat{\theta}) $
\begin{equation}
\widehat{\cal A}_z(\hat{\rho}, \hat{\theta} ) = \widehat{\cal A}_z' (\hat{\rho}, \hat{\theta}) + \widehat{\cal A}_z'' (\hat{\rho},\hat{\theta}) \,.
\label{eq:TotField}
\end{equation}

\subsubsection{Particular Solution $\widehat{\cal A}_z''$}
The particular solution of \pref{eq:PDE2} at an arbitrary point inside the hole is given by \cite{Bal89}
\begin{equation}
\widehat{\cal A}_{z}''(\hat{\rho},\hat{\theta})  = -\hat{\mu} \sum_{q=1}^{P} \int_0^{2\pi} J_s^{(q)}(\theta_q') \widehat{G}\left( \widehat{\vect{r}}(\hat{\rho}, \hat{\theta}), \vect{r}_q(\theta_q')\right) a_q\, d\theta_q'\,.
\label{eq:ParSol1}
\end{equation}
This formula is the superposition of the potential caused by the equivalent current introduced on each conductor.
The integral kernel in~\pref{eq:ParSol1} reads
\begin{equation}
\widehat{G}\left( \vect{r}, \vect{r}'\right) = \frac{j}{4} {\cal H}_0^{(2)} \left( \hat{k} \abs{ \vect{r} - \vect{r}'} \right) \,
\label{eq:Ghat}
\end{equation}
and corresponds to the Green's function of a homogeneous\footnote{We remark that we are solving~\pref{eq:PDE2} only inside the contour $\hat{c}$ shown in the right panel of Fig.~\ref{fig:replaceconfig}. In this region, the medium is homogeneous.} medium~\cite{Bal89} with permittivity $\hat{\varepsilon}$ and permeability $\hat{\mu}$.

\subsubsection{General Solution $\widehat{\cal A}_z'$}
The general solution of~\pref{eq:PDE2} is given by~\cite{Bal89}
\begin{equation}
\widehat{\cal A}_{z}'(\hat{\rho}, \hat{\theta})  = \sum_{n=-\widehat{N}}^{\widehat{N}} C_n {\cal J}_{\abs{n}}\left( \hat{k} \hat{\rho} \right) \operatorname{e}^{jn\hat{\theta}} \,,
\label{eq:GenSol1}
\end{equation}
where coefficients $C_n$ are found by enforcing the boundary condition~\pref{eq:BC1}, and are stored into a vector $\vect{C} = \begin{bmatrix} C_{-\widehat{N}} & \hdots & C_{\widehat{N}} \end{bmatrix}^T$. By substituting~\pref{eq:ParSol1} and~\pref{eq:GenSol1} into~\pref{eq:TotField}, the boundary condition~\pref{eq:BC1} can be imposed using the method of moments~\cite{Wal08}, a mathematical technique to solve integral equations numerically. This process, which is analogous to the one detailed in~\cite{TPWRD1}, provides an algebraic expression for the coefficients $\vect{C}$
\begin{equation}
\vect{C} = \matr{D}_1 \left( \widehat{\vect{A}}  + \hat{\mu} \widehat{\matr{G}}_0 \vect{J} \right) \,,
\label{eq:Cvalue1}
\end{equation}
where $\matr{D}_1$ is a diagonal matrix with diagonal entries $(n,n)$ equal to
\begin{equation}
\left[ \matr{D}_1\right]_{[n,n]} =  ({\cal J}_{\abs{n}}(\hat{k} \hat{a}))^{-1}
\end{equation}
for $n = -\widehat{N}, \hdots, \widehat{N}$. The matrix $\widehat{\matr{G}}_0$ in~\pref{eq:Cvalue1} is the discrete counterpart of the Green's function~\pref{eq:Ghat} and can be obtained with the procedure presented in~\cite{TPWRD1}.

\subsubsection{Vector potential $\widebar{\cal A}_z$}

We calculate the term $\widebar{\cal A}_z(\hat{\rho},\hat{\theta})$ in~\pref{eq:Jhateq1}, which is the fictitious field inside the hole 
when the hole and equivalent currents inside it are replaced by the ground medium. This term is the solution of the Helmholtz equation~\pref{eq:PDE2} with the right hand side term set to zero, and $\hat{k}$ replaced by the wavenumber $k_g = \sqrt{\omega \mu_0 \left( \omega \varepsilon_0 - j \sigma_g \right)}$ of the surrounding ground. Accounting for the boundary condition~\pref{eq:BC1}, the solution is given by~\cite{Bal89}
\begin{equation}
\widebar{\cal A}_z( \hat{\rho}, \hat{\theta}) = \sum_{n=-\widehat{N}}^{\widehat{N}} \widehat{A}_n \frac{{\cal J}_{\abs{n}} (k_g \hat{\rho}) }{ {\cal J}_{\abs{n}}(k_g \hat{a})} \operatorname{e}^{jn\hat{\theta}}\,.
\label{eq:RepField1}
\end{equation}

\subsubsection{Equivalent Hole Current}
We can finally derive the equivalent current $\widehat{J}_s(\hat{\theta})$. We substitute~\pref{eq:JHat1}, \pref{eq:RepField1}, \pref{eq:GenSol1}, and~\pref{eq:ParSol1}  into \pref{eq:Jhateq1} to get the equation
\begin{align}
 \sum_{n=-\widehat{N}}^{\widehat{N}}&  \frac{\widehat{J}_n}{2\pi \hat{a}}  \operatorname{e}^{jn\hat{\theta}}  = \left[  \sum_{q=1}^{P} \int_0^{2\pi}   J_s^{(q)}(\theta_q') \frac{\partial \widehat{G}}{\partial \hat{\rho}} a_q d\theta_q'  \right.\label{eq:JHat2} \\
& \left. {} -  \sum_{n=-\widehat{N}}^{\widehat{N}} \left( \frac{\hat{k} C_n}{\hat{\mu}}  {\cal J}_{\abs{n}}'\left( \hat{k} \hat{a} \right) - \widehat{A}_n \frac{k_g {\cal J}_{\abs{n}} '(k_g \hat{a}) }{ \mu_0{\cal J}_{\abs{n}}(k_g \hat{a})} \right) \operatorname{e}^{jn \hat{\theta}}  
\right] \,.
\nonumber
\end{align}
The obtained integral equation is solved for the coefficients $\widehat{J}_n$ using the method of moments~\cite{Wal08} to obtain, with a process similar to the one given in~\cite{TPWRD1}, the following formula
\begin{equation}
\widehat{\vect{J}} = \widehat{\matr{Y}}_s \widehat{\vect{A}} + \vect{T} \vect{J} \,.
\label{eq:surf2}
\end{equation}
Equation~\pref{eq:surf2} shows that the equivalent current $\widehat{\vect{J}}$ that represents the cable-hole system is made by two components. The first term is the contribution of an empty hole without conductors inside. This term is analogous to the surface admittance~\pref{eq:surf1} of a single round conductor~\cite{TPWRD1,DeZ05}, and involves the diagonal matrix $\widehat{\vect{Y}}_s$ with entries
\begin{align}
\left[ \widehat{\matr{Y}}_s\right]_{[n,n]} &= 2\pi \hat{a} \left[ \frac{k_g}{\mu_0} \frac{{\cal J}_{\abs{n}} '(k_g \hat{a}) }{ {\cal J}_{\abs{n}}(k_g \hat{a})}  - \frac{\hat{k}}{\hat{\mu}}\frac{{\cal J}_{\abs{n}} '(\hat{k} \hat{a}) }{ {\cal J}_{\abs{n}}(\hat{k} \hat{a})} \right ]\,,
\end{align}
for $n = -\widehat{N}, \hdots, \widehat{N}$. The second term in~\pref{eq:surf2} is due to the conductors present in the hole. The transformation matrix $\vect{T}$ maps the currents $\vect{J}$ on the conductor boundaries to the equivalent current  $\widehat{J}_s(\hat{\theta})$ on the hole boundary. The transformation matrix $\matr{T}$ is given by
\begin{equation}
\matr{T} = 2\pi \hat{a} \left[ \widetilde{\matr{G}}_0 - \matr{D}_2 \widehat{\matr{G}}_0 \right] \,,
\end{equation}
where $\matr{D}_2$ is a diagonal matrix with entries
\begin{equation}
\left[ \matr{D}_2\right]_{[n,n]} = \hat{k} \frac{{\cal J}_{\abs{n}} '(\hat{k} \hat{a}) }{ {\cal J}_{\abs{n}}( \hat{k} \hat{a})} \,,
\end{equation}
for $n= -\widehat{N}, \hdots, \widehat{N}$, and
matrix $\widetilde{\matr{G}}_0$ comes from the discretization of the derivative of the Green's function~\pref{eq:Ghat}. 
Expression~\pref{eq:surf2} is one of the main contributions of this work, since it provides an efficient way to represent the cable-hole system, which in turn will enable a fast computation of the cable impedance.

\section{Inclusion of Ground Return Effects}
\label{sec:GroundReturn}

At this point, we have replaced the cable-hole system with a single equivalent current placed on the hole boundary, as shown in the right panel of Figure~\ref{fig:replaceconfig}.  We now couple the cable-hole representation with an integral equation describing the behavior of the air-ground medium which surrounds the hole. This will allow us to determine the magnetic vector potential $\widehat{{\cal A}}_z$ on the hole boundary and then, in Sec.~\ref{sec:efie}, calculate the p.u.l. impedance of the cable.

By the definition of magnetic vector potential, we can relate the current and vector potential through the integral equation~\cite{Har61}
\begin{equation}
\widehat{{\cal A}}_z(\hat{a}, \hat{\theta}) = - \mu_0 \int_0^{2\pi} \widehat{J}_s(\hat{\theta} ') {G}_{g}\left(\widehat{\vect{r}}(\hat{a}, \hat{\theta}),  \widehat{\vect{r}}(\hat{a}, \hat{\theta}')\right) \hat{a} d \hat{\theta}' \,,
\label{eq:outerefie}
\end{equation}
where $G_g$ is the Green's function of medium made by two layers, in our case air and ground. This Green's function reads~\cite{yang96}
\begin{align}
{G}_g&(x,y, x', y') =\frac{1}{4\pi} \int_{-\infty}^{\infty} \frac{\operatorname{e}^{ -j\beta_x \left( x - x' \right) }}{\sqrt{\beta_x^2 - k_{g}^2 } } \,   \label{eq:GreenUnderground} \\ 
\,  &\left[ \operatorname{e}^{ - \abs{y - y'}\sqrt{\beta_x^2 - k_{g}^2} } 
+ R_{TM} \operatorname{e}^{ (y + y') \sqrt{\beta_x^2 - k_{g}^2} }  \right ] \, d\beta_x \,,
\nonumber
\end{align}
where
\begin{equation}
R_{TM} = \frac{\sqrt{\beta_x^2 - k_{g}^2 }     -  \sqrt{\beta_x^2 - k_{0}^2 }   }{\sqrt{\beta_x^2 - k_{g}^2 }    + \sqrt{\beta_x^2 - k_{0}^2 }  } \,,
\end{equation}
where $k_0 = \omega \sqrt{\mu_0 \varepsilon_0}$ is the wavenumber of air.
In~\pref{eq:GreenUnderground}, we use $x$, $y$, $x'$ and $y'$ to express the x-component and y-component of the position vectors  $\widehat{\vect{r}}(\hat{a}, \hat{\theta})$ and $\widehat{\vect{r}}(\hat{a}, \hat{\theta} ')$.
We next substitute the truncated Fourier expansions~\pref{eq:BC1} and~\pref{eq:JHat1} into~\pref{eq:outerefie}, and apply the method of moments~\cite{Wal08} to convert the resulting integral equation into a standard algebraic equation
\begin{equation}
\widehat{\vect{A}} = - \mu_0 \matr{G}_g \widehat{\vect{J}} \,,
\label{eq:EFIEhalf}
\end{equation}
where $\matr{G}_g$ is the discretization of the Green's function~\pref{eq:GreenUnderground}. 
By substituting~\pref{eq:surf2} into~\pref{eq:EFIEhalf} we obtain the coefficients of the magnetic vector potential on the hole boundary
\begin{equation}
\widehat{\vect{A}} = - \mu_0 \left(\vect{1} + \mu_0 \matr{G}_g \widehat{\matr{Y}}_s \right)^{-1} \matr{G}_g \matr{T} \vect{J} \,,
\label{eq:Arel2}
\end{equation}
where $\matr{1}$ is the identity matrix.
\section{Computation of Per-unit-length Parameters}
\label{sec:efie}

In order to compute the p.u.l. impedance of the cable, we need the electric field on the boundary $c_p$ of each conductor, which can be obtained from the vector potential $\widehat{\cal A}_z$ and the scalar potential $V$ as~\cite{Har61}
\begin{equation}
E_z (\vect{r}_p(\theta_p)) = - j\omega \widehat{\cal A}_z - \frac{\partial V}{\partial z}\,.
\label{eq:efie1}
\end{equation}
Next, we substitute~\pref{eq:Tel1} and~\pref{eq:TotField} into~\pref{eq:efie1} to obtain, after discretization with the method of moments~\cite{Wal08}, the algebraic equation
\begin{equation}
\vect{E} = -j\omega \matr{\widehat{H}} \vect{C} + j\omega \hat{\mu} \widehat{\matr{G}}_c \vect{J} + \matr{U} \left[\ppulpm{R}(\omega) + j \omega \ppulpm{L}(\omega) \right] \matr{U}^T\matr{J} \,,
\label{eq:efie2}
\end{equation}
where matrices $\widehat{\matr{G}}_c$ and $\widehat{\matr{H}}$ come from the discretization of particular solution~\pref{eq:ParSol1} and general solution~\pref{eq:GenSol1}, respectively. The constant matrix $\matr{U}$ is the same as the one defined in~\cite{TPWRD1}.
By substituting~\pref{eq:Cvalue1} and~\pref{eq:Arel2} into~\pref{eq:efie2}, we get
\begin{equation}
\vect{E} = j\omega \matr{\Psi} \vect{J}  + \matr{U} \left[ \ppulpm{R}(\omega) + j \omega \ppulpm{L}(\omega) \right] \matr{U}^T \matr{J} \,,
\label{eq:efie3}
\end{equation}
where
\begin{equation}
\matr{\Psi} =  \widehat{\matr{H}} \matr{D}_1  \left[ {\mu}_0 \left(\vect{1} + {\mu}_0 \matr{G}_g \widehat{\matr{Y}}_s \right)^{-1} \matr{G}_g \matr{T}  - \hat{\mu} \widehat{\matr{G}}_0 \right] + \hat{\mu} \widehat{\matr{G}}_c  \,. 
\label{eq:Psi}
\end{equation}
In~\pref{eq:Psi}, the term between square brackets accounts for the presence of hole and of the air-ground interface and was not considered in previous works~\cite{TPWRD1,TPWRD2}, which only accounted for the factor $\hat{\mu} \widehat{\matr{G}}_c$. This last term is the only one needed to model a cable which is buried into a homogeneous soil at infinite depth and is not surrounded by a hole. From~\pref{eq:efie3}, the p.u.l. resistance and inductance matrices can be finally obtained using the steps presented in~\cite{TPWRD1}
\begin{align}
\ppulpm{R}(\omega)&= \Re { \left( \matr{U}^T\left( \matr{1} - j \omega \matr{Y}_s \Psi  \right) ^{-1} \matr{Y}_s \vect{U} \right)^{-1}} \,, \label{eq:R} \\
\ppulpm{L}(\omega) &= \omega^{-1}\Im { \left( \matr{U}^T\left( \matr{1} - j \omega\matr{Y}_s \Psi  \right) ^{-1} \matr{Y}_s \vect{U} \right)^{-1}} \label{eq:L} \,.
\end{align}
Expressions~\pref{eq:R} and~\pref{eq:L} are used in the numerical examples of Sec.~\ref{sec:Results} to calculate the p.u.l. parameters of several cable systems.

\section{Extension to Hollow Conductors and Multiple Holes}
\label{sec:MultipleHole}

For the sake of clarity, we have described the proposed method considering only solid round conductors buried into a single hole. However, the proposed technique can handle any arrangement of solid and hollow conductors buried into multiple holes dug in lossy soil. In this section, we discuss how hollow conductors and multiple holes can be easily introduced in the theoretical frameworks discussed so far.

In order to include a hollow conductor, we first replace it with two equivalent currents placed on the inner and outer boundary of the conductor~\cite{TPWRD2}. Then, the surface admittance operator of a hollow conductor~\cite{TPWRD2} provides the relation between equivalent currents and electric field necessary to form~\pref{eq:surf1}.
In presence of multiple holes, the process of Sec.~\ref{sec:SAcablehole} is first applied to each hole independently. An equivalent current~\pref{eq:JHat1} is introduced on the boundary of the hole, and related to the equivalent currents present inside that specific hole through~\pref{eq:surf2}. Then, one integral per hole is added to the right hand side of~\pref{eq:outerefie}.

\section{Numerical Results}
\label{sec:Results}

\subsection{Three Single Core Cables Buried in Earth}
\label{sec:Case1}

\begin{figure}[t]
\begin{center}
\begin{tikzpicture}[scale=0.25, every node/.style={scale=0.25}]

\tikzstyle{line} = [draw, -latex', <->];
\draw [fill=black!20] (0,0.5) rectangle (16,8);
\draw [fill=white] (0,6.8) rectangle (16,8);

\draw [fill=white]  (2.5,3.5) ellipse (2 and 2);
\draw [fill=black!80] (2.5,3.5) ellipse (1.8 and 1.8);
\draw [fill=white] (2.5,3.5) ellipse (1.7 and 1.7);
\draw [fill=black!50] (2.5,3.5) ellipse (1.3 and 1.3);

\draw [fill=white]  (7.0,3.5) ellipse (2 and 2);
\draw [fill=black!80] (7.0,3.5) ellipse (1.8 and 1.8);
\draw [fill=white] (7.0,3.5) ellipse (1.7 and 1.7);
\draw [fill=black!50] (7.0,3.5) ellipse (1.3 and 1.3);

\draw [fill=white]  (11.5,3.5) ellipse (2 and 2);
\draw [fill=black!80] (11.5,3.5) ellipse (1.8 and 1.8);
\draw [fill=white] (11.5,3.5) ellipse (1.7 and 1.7);
\draw [fill=black!50] (11.5,3.5) ellipse (1.3 and 1.3);

\node (p1) at (14,3.5) {};
\node (p2) at (14,6.8) {};
\path [line] (p1) -- (p2) {};

\node (p3) at (2.5,5.8) {};
\node (p4) at (7.0,5.8) {};
\path [line] (p3) -- (p4)  {};

\node (p5) at (4.75,5.8) [above]{\huge spacing};
\node (p6) at (14,5.0) [right] {\huge depth};

\node at (12.0,0.95) {\huge ground ($\varepsilon_0, \mu_0, \sigma_g$)};

\node at (12.5,7.5) [right] {\huge air ($\varepsilon_0 , \mu_0$)};
\end{tikzpicture}
\caption{System of three single core cables used for validation in Sec.~\ref{sec:Results}. Conductive media are shown in gray while insulating media are shown in white. }
\label{fig:Validation1}
\end{center}
\end{figure}
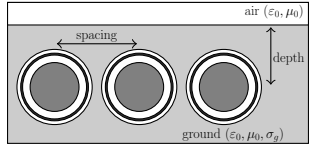

We compare the proposed MoM-SO method against a commercial FEM solver (COMSOL Multiphysics~\cite{COMSOL}) and the ``cable constant'' formulas~\cite{Ame80}. As a first test case, we consider a system of three single core (SC) cables buried in ground at a depth of $1~{\rm m}$, as shown in Fig.~\ref{fig:Validation1}. With this example, we also demonstrate that MoM-SO can handle multiple holes and hollow conductors.

\subsubsection{Geometrical and Material Properties}

The geometrical and material parameters of the three SC cables are presented in Table~\ref{tab:parameters}.  Two different values for cable spacing are considered: $s = 2~{\rm m}$ and $s = 85~{\rm mm}$. The conductivity of ground is set to $0.01 ~{\rm S/m}$.

\begin{table}[t]
\centering
\caption{Single core cables of Sec.~\ref{sec:Case1}: geometrical and material parameters}
\begin{tabular}{|c|c|}
\hline
Core & Outer diameter = 39~mm, $\rho = 3.365\cdot 10^{-8} {\rm~\Omega \cdot m}$ \\ \hline
Insulation & $t = 18.25 {\rm~mm} $, $\epsilon_r = 2.85$ \\ \hline
Sheath & $t = 0.22{\rm~mm}$, $\rho = 1.718\cdot 10^{-8} {\rm~\Omega \cdot m}$ \\ \hline
Jacket & $t = 4.53{\rm~mm}$, $\epsilon_r = 2.51$ \label{tab:SCparameter} \\ \hline
\end{tabular}
\label{tab:parameters}
\end{table}

\subsubsection{Simulation Setup}

Both FEM and MoM-SO are set up to extract the impedance matrix of the system of six conductors (three core conductors plus three hollow screens), assuming the return path for the currents to be at infinity. Impedance is evaluated at 31 frequency points logarithmically spaced between 1~Hz and 1~MHz.  

In MoM-SO, we set to 4 the order $N_p$ and $\widehat{N}$ of the Fourier expansions~\pref{eq:Eexp1}, \pref{eq:Jexpan1}, \pref{eq:BC1},  and~\pref{eq:JHat1}. This value is sufficient to accurately describe proximity effects even when the SC cables are close to each other~\cite{spi2013}. In the FEM solver, the solution mesh has to be carefully set up to achieve good accuracy. Ground has to be meshed up to a distance of three times the skin depth, in order to properly calculate  ground return current. For the first 25 frequency points, we used a mesh with 725,020 triangles for the $s=85\,{\rm mm}$ case, and 837,618 triangles for the $s=2\,{\rm m}$ case. 
At the last six frequency points, which are spread between 100 kHz and 1 MHz, skin depth becomes extremely small, and the mesh has to be refined inside the conductors. This required the use of the so-called boundary layer elements, and increased mesh size to 1,053,638 for the $s = 2\,{\rm m}$ case.

\subsubsection{Continuously-grounded Screens}

\begin{figure}[t]
\centering
\includegraphics[scale=0.75]{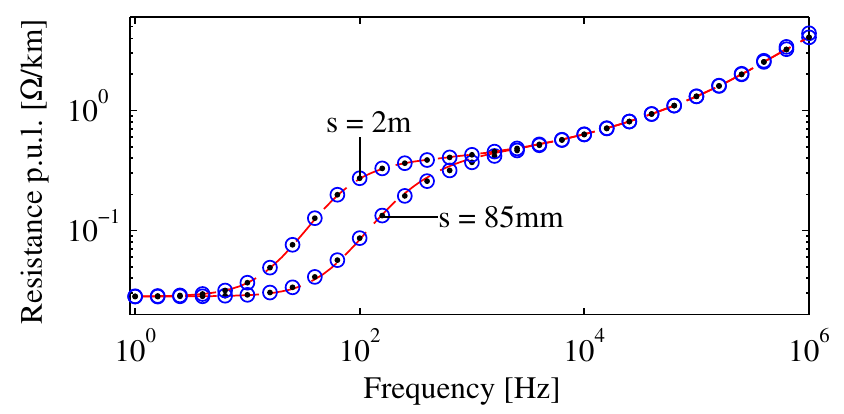}
\includegraphics[scale=0.75]{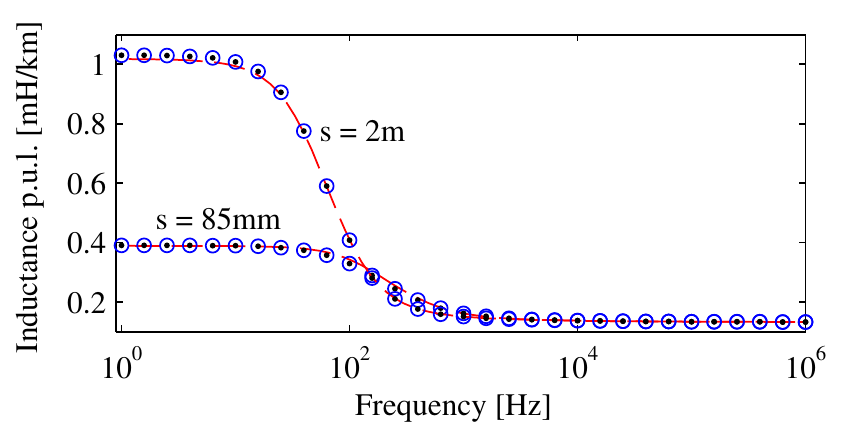}
\caption{Cable system of Sec.~\ref{sec:Case1}: positive-sequence resistance (top panel) and inductance (bottom panel) computed using FEM ({\color{blue} ${\bf \mathlarger \circ}$}), MoM-SO ($\oldcdot[-1.5pt]$), and cable constant ({\color{red} -\,-}). Screens are continuously grounded.}
\label{fig:positiveseq_gnd_case1ab}
\end{figure}

\begin{figure}[t]
\centering
\includegraphics[scale=0.75]{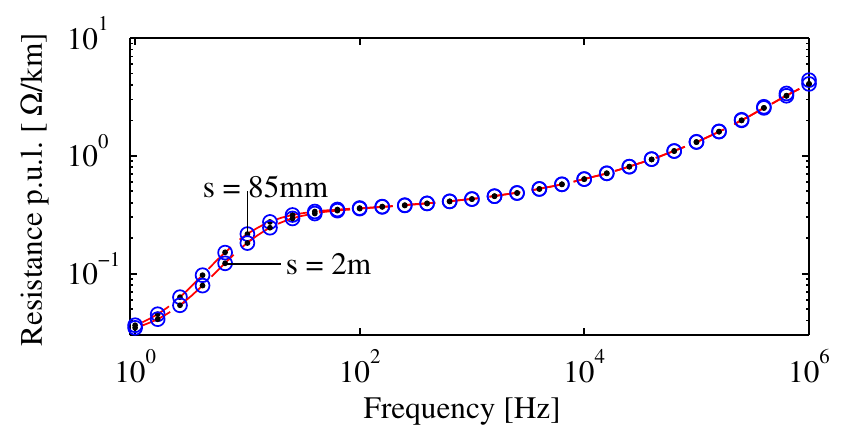}
\includegraphics[scale=0.75]{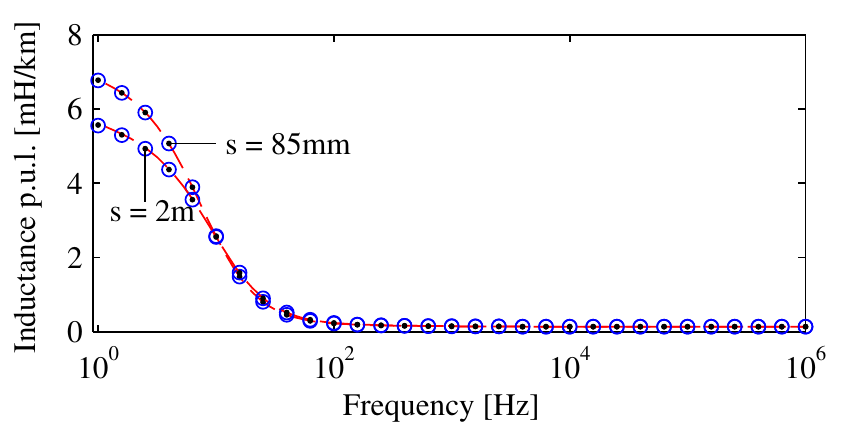}
\caption{As in Fig.~\ref{fig:positiveseq_gnd_case1ab}, but when a zero-sequence is applied to the cable.}
\label{fig:zeroseq_gnd_case1ab}
\end{figure}

\addedtwo{We consider two different scenarios for this example: grounded screens and open screens. In the first case, we assume ideal grounding, and we calculate the $3 \times 3$ impedance matrix of the cable from the $6\times 6$ impedance matrix \added{by setting the potentials of the screens to zero}.}
The positive-sequence resistance and inductance obtained with MoM-SO, FEM and cable constant formulas are presented in Fig.~\ref{fig:positiveseq_gnd_case1ab}. The zero-sequence resistance and inductance are instead shown in Fig.~\ref{fig:zeroseq_gnd_case1ab}.\footnote{ \added{ Positive-sequence impedance is defined as the ratio of positive-sequence voltages and currents. Similarly, zero-sequence impedance is defined as the ratio of zero-sequence voltages and currents~\cite{For18}.}}
The excellent agreement observed between FEM and MoM-SO validates the proposed technique. Since screens are grounded, there is little proximity effect between the three SC cables. Hence, cable constant formulas provide accurate results. \added{The p.u.l. resistance is different for $s=2~{\rm m}$ and $s= 85~{\rm mm}$ because  mutual   impedance is different in the two cases.}
\addedtwo{We remark that the ideal grounding assumption has been used here only for simplicity. Such assumption is not required by the proposed MoM-SO method, that can be used to study more complex cable systems with cross-bonding, as shown in~\cite{TPWRD2}.}

\subsubsection{Open Screens}
\begin{figure}[t]
\centering
\includegraphics[scale=0.75]{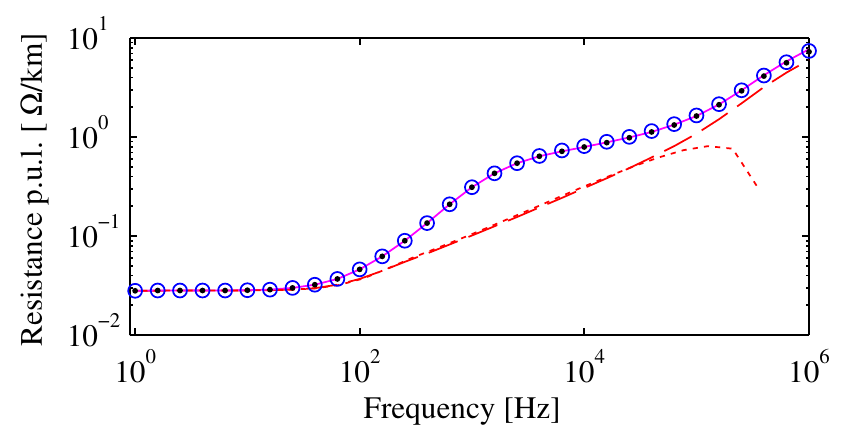}
\includegraphics[scale=0.75]{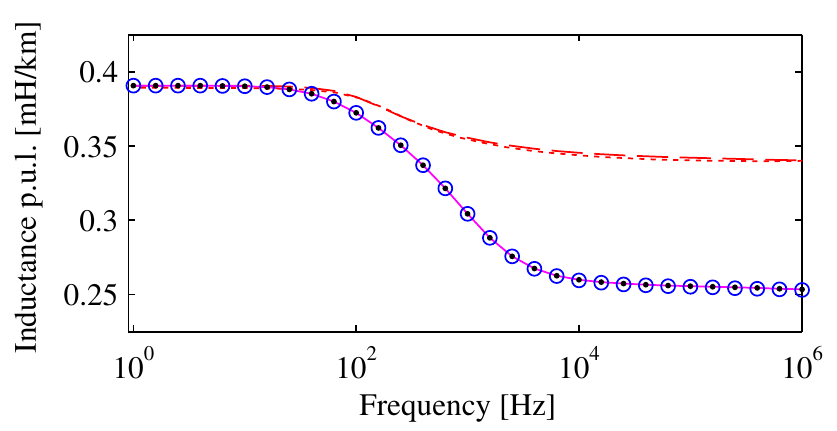}
\caption{Cable system of Sec.~\ref{sec:Case1}: positive-sequence resistance (top panel) and inductance (bottom panel) computed using FEM ({\color{blue} ${\bf \mathlarger \circ}$}), MoM-SO ($\oldcdot[-1.5pt]$), cable constant with Pollaczek ground return ({\color{red} \xdash[0.5em] \xdash[0.5em]}), cable constant with Saad ground return ({\color{red} \xdash[0.1em]  \xdash[0.1em] \xdash[0.1em] \xdash[0.1em]}) and
MoM-SO with approximate ground return effects~\cite{TPWRD2} ({\color{magenta} \xdash[1.2em]}). The screens of the cables are open.}
\label{fig:positiveseq_open_case1ab}
\end{figure}

\addedtwo{In this second case, screens are not grounded but left open. As a consequence, large sheath overvoltages~\cite{Nag83, Gus95} and a significant proximity effect between the three SC cables can develop.} \added{When screens are left open, screen currents are zero, which allows us to reduce the $6\times 6$ matrix to a $3\times3$ matrix.
}
Figure~\ref{fig:positiveseq_open_case1ab} shows the positive-sequence resistance and inductance for the case where cables are close together ($s=85\,{\rm mm}$).  MoM-SO and FEM accurately capture the impedance variation due to skin and proximity effect in conductors and ground. Cable constant formulas \added{with Pollaczek and Saad ground return formulas}, on the other hand, return accurate results only at low frequency, and become inaccurate beyond 100~Hz. 
\added{Moreover, Saad formula~\cite{Saad96} returns a negative resistance at high frequency. If cable spacing is increased to 2~m, the results from cable constant formulas agree reasonably with FEM and MoM-SO, confirming that the deviation observed in Fig.~\ref{fig:positiveseq_open_case1ab} is due to proximity effects.}
 Figure~\ref{fig:positiveseq_open_case1ab} also shows the resistance and inductance obtained with our previous method~\cite{TPWRD2}, where MoM-SO is used to model proximity effects in conductors, and cable constant formulas \added{(Pollaczek)} are used to model ground return effects.

\subsubsection{Timing Results}

Table~\ref{tab:Timing1} shows the CPU time taken by MoM-SO and FEM to analyze the cable system. FEM requires more than 6~minutes per frequency point, while MoM-SO only 0.8~s. This dramatic speed up, beyond 400X, comes from the fact that, with the MoM-SO method, one has to mesh neither the cross section of the conductors nor the surrounding ground where return current may flow. On the other hand, the complex mesh needed to capture ground return effects and skin effect at high frequency makes FEM very time consuming. Moreover, with FEM, the user must spend extra time to properly set up the mesh generator, since default settings may not lead to accurate results. MoM-SO, instead, being meshless, is much easier to use, and can be fully automated~\cite{spi2013}.

\begin{table}[t]
\centering
\caption{Example of Sec.~\ref{sec:Case1}: CPU time required to compute the impedance at one frequency}
\begin{tabular}{|c|c|c|c|}
\hline
Case & MoM-SO (Proposed)  & FEM  & Speed-up\\  \hline
$s=85 {\rm mm}$ & 0.80 s & 371.21 s & 464 X \\ \hline
$s=2 {\rm m}$ & 0.80 s &  452.77 s &  566 X\\ \hline
\end{tabular} 
\label{tab:Timing1} 
\end{table}

\subsection{Effect of Ground Resistivity}
\label{sec:results_case1c}

\begin{figure}[t]
\centering
\includegraphics[scale=0.75]{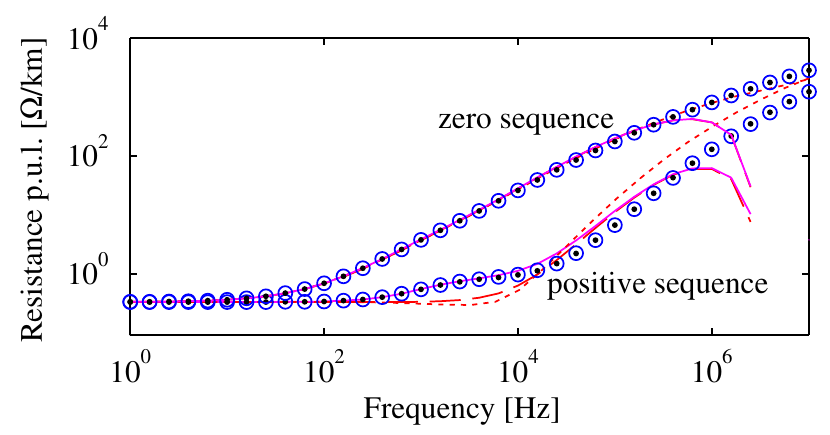}
\includegraphics[scale=0.75]{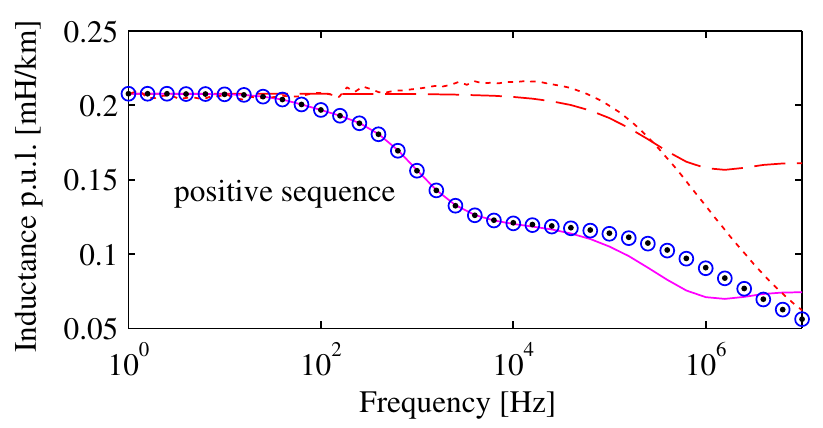}
\caption{Cable system considered in Sec.~\ref{sec:results_case1c}: resistance (top panel) and inductance (bottom panel) computed using FEM ({\color{blue} ${\bf \mathlarger \circ}$}), MoM-SO ($\oldcdot[-1.5pt]$), cable constant with Pollaczek ground return ({\color{red} \xdash[0.5em] \xdash[0.5em]}),  cable constant with Saad ground return ({\color{red} \xdash[0.1em]  \xdash[0.1em] \xdash[0.1em] \xdash[0.1em]}), and MoM-SO with approximate ground return effects \cite{TPWRD2} ({\color{magenta} \xdash[1.2em]}). Phase conductors are open, and current is injected in the sheaths.}
\label{fig:Results_case1c}
\end{figure}

We consider the three SC cables with spacing $s=85\,{\rm mm}$ and ground conductivity $\sigma_g = 100~{\rm S/m}$. This high conductivity value is used to show how proximity effects in ground influence the cable impedance. We let the phase conductors open and inject currents in the sheaths. 
Figure~\ref{fig:Results_case1c} shows the resistance and inductance obtained in this scenario with MoM-SO, FEM and the method of~\cite{TPWRD2}, which neglects proximity effects in ground.  The excellent agreement between MoM-SO and FEM shows that the proposed method correctly captures proximity effects in both conductors and ground. \added{Proximity effects inside conductors start being relevant at 100~Hz. Proximity effects in ground develop instead above 10~kHz, as can be seen by comparing the results from the proposed technique against those computed with the method of~\cite{TPWRD2}, which neglects proximity in ground. This hybrid method uses the MoM-SO approach for  conductors, and Pollaczek formula for ground effects. Since for this configuration Pollaczek formula returns a negative resistance above 2~MHz, the corresponding curve and the curve of~\cite{TPWRD2} are truncated.}

\subsection{Three Single-Core Cables  Inside a Tunnel}
\label{sec:results_case2}

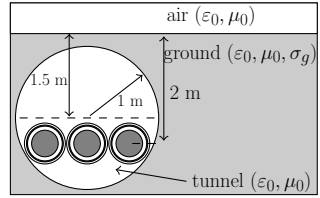
\begin{figure}[t]
\begin{center}
\begin{tikzpicture}[scale=0.34, every node/.style={scale=0.31}]

\tikzstyle{line} = [draw, very thick, -latex', <->];
\draw [fill=black!20] (4,0.5) rectangle (16,8);
\draw [fill=white] (4,6.8) rectangle (16,8);

\draw [fill=white]  (7,3.5) ellipse (2.8 and 2.8);

\draw [fill=white]  (5.35,2.5) ellipse (2/2.5 and 2/2.5);
\draw [fill=black!80] (5.35,2.5) ellipse (1.8/2.5 and 1.8/2.5);
\draw [fill=white] (5.35,2.5) ellipse (1.7/2.5 and 1.7/2.5);
\draw [fill=black!50] (5.35,2.5) ellipse (1.3/2.5 and 1.3/2.5);

\draw [fill=white]  (7.0,2.5) ellipse (2/2.5 and 2/2.5);
\draw [fill=black!80] (7.0,2.5) ellipse (1.8/2.5 and 1.8/2.5);
\draw [fill=white] (7.0,2.5) ellipse (1.7/2.5 and 1.7/2.5);
\draw [fill=black!50] (7.0,2.5) ellipse (1.3/2.5 and 1.3/2.5);

\draw [fill=white]  (8.65,2.5) ellipse (2/2.5 and 2/2.5);
\draw [fill=black!80] (8.65,2.5) ellipse (1.8/2.5 and 1.8/2.5);
\draw [fill=white] (8.65,2.5) ellipse (1.7/2.5 and 1.7/2.5);
\draw [fill=black!50] (8.65,2.5) node (v1) {} ellipse (1.3/2.5 and 1.3/2.5);

\node (p1) at (10,2.5) {};
\node (p2) at (10,6.8) {};
\node at (10,4.5)  [right] {\Huge \,2 m};

\draw [<->] (p1) -- (p2) {};

\node (p3) at (7,3.5) {};
\node (p4) at (7.3 + 2.8/1.414 ,3.5 + 2.5/1.414) {};
\node at (7.7 + 1.4/1.414, 3.5 + 1.4/1.414) [below] {\huge 1 m};
\draw [->] (p3) -- (p4)  {};


\node at (13.0,6.0) {\Huge ground ($\varepsilon_0, \mu_0, \sigma_g$)};

\node at (10,7.5) [right] {\Huge air ($\varepsilon_0 , \mu_0$)};
\draw [dashed]  (v1) edge (p1);
\node at (11,1) [right] {\Huge tunnel ($\varepsilon_0 , \mu_0$)};
\node (p6) at (11,1) {};
\node (p7) at (8,1.5) {};
\draw [->] (p6) -- (p7) {};
\node (v2) at (6.3,6.9) {};
\node (v5) at (6.3,3.45){};
\draw  [<->](v5) edge (v2);
\node at (6.37,5) [left]{\huge 1.5 m};
\node(p15) at (4.25,3.5) {};
\node(p16) at (9.85,3.5) {};
\draw [dashed] (p15) -- (p16) {};
\end{tikzpicture}
\caption{System of three single-core cables in a tunnel considered in Sec.~\ref{sec:results_case2}. Conductive media are shown in gray while insulating media are shown in white. }
\label{fig:Validation2}
\end{center}
\end{figure}

Finally, we consider a system of three SC cables placed inside a tunnel. The cross-section of the system is depicted in Fig.~\ref{fig:Validation2}. Cables are spaced by $s=85~{\rm mm}$, and their characteristics are reported in Table~\ref{tab:SCparameter}. Sheaths are left open at both ends.

\begin{figure}[t]
\centering
\includegraphics[scale=0.75]{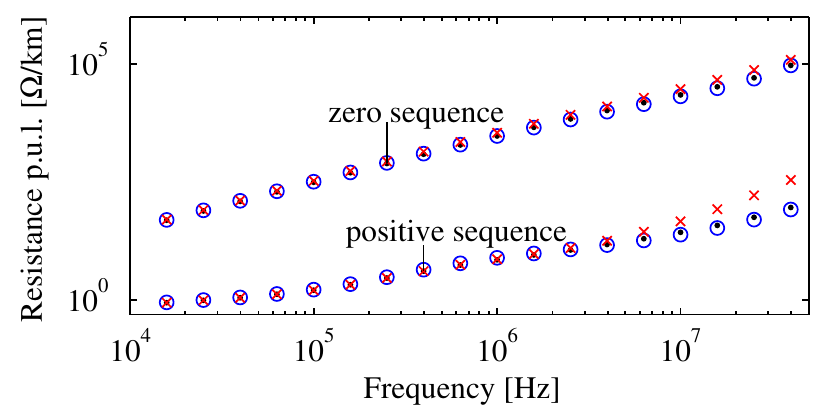}
\includegraphics[scale=0.75]{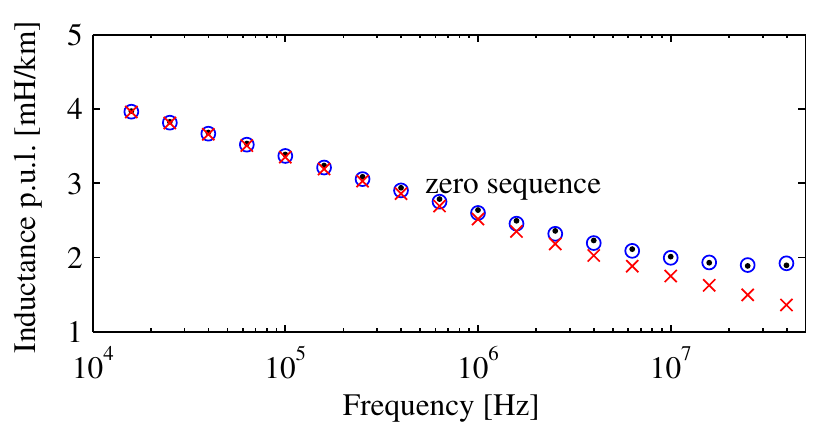}
\caption{System of three SC cables in a tunnel considered in Sec.~\ref{sec:results_case2}: resistance (top panel), and inductance (bottom panel) computed with FEM ({\color{blue} ${\bf \mathlarger \circ}$}) and MoM-SO ($\oldcdot[-1.5pt]$). In order to show the effect of the tunnel, the resistance and inductance of the cables buried directly in ground are also shown ({\color{red} $\times$}).}
\label{fig:Results_case2}
\end{figure}

Firstly, FEM and MoM-SO are used to compute the positive- and zero-sequence impedance of the cable in presence of the tunnel. 
Secondly, the computation is repeated with the tunnel removed and the cables buried directly in ground. 
The resistance and inductance values obtained for both cases are shown in Fig.~\ref{fig:Results_case2}. The influence of the tunnel on the cable impedance is visible above 3~MHz on both resistance and inductance. The results obtained with MoM-SO match closely those obtained with FEM.
 However, MoM-SO took only 0.29~s per frequency point against the 498.3~s taken by FEM, for a speed up of 1,734 times. The high computational efficiency of MoM-SO makes it practical for routine use, differently from FEM which can be quite time-consuming and requires special care in the setup of the mesh.

\section{Conclusion}

This paper presents MoM-SO, an efficient numerical technique to compute the series resistance and inductance of power cables while accounting for skin, proximity and ground return effects. MoM-SO can handle any arrangement of solid and tubular round conductors buried in a lossy ground medium. Conductors can be placed in one or more holes or tunnels excavated in ground.  MoM-SO accounts for several factors that influence cable impedance, namely skin effect, proximity effects in both conductors and ground, ground return current, finite burial depth, and the presence of a hole or tunnel around the cable. Comparison against finite elements shows that MoM-SO accurately predicts such phenomena from the Hz to the MHz range. MoM-SO is considerably faster than finite elements, since speed-ups beyond 1000X have been demonstrated. Also, since MoM-SO avoids mesh-related issues, it is easier to use than finite elements. In conclusion, MoM-SO makes the modeling of power cables for transient analyses simpler and more accurate, especially in those scenarios where proximity effects cannot be neglected and, consequently, widely-used analytic formulas cannot be applied~\cite{TPWRD2,ipst2013}.

\section{Acknowledgement}
Authors thank Dr. Bj\o rn Gustavsen from SINTEF Energy Research, Norway, for providing the test cases of Sec.~\ref{sec:Results}.

\bibliography{IEEEabrv,biblio}
\bibliographystyle{IEEEtran}

\vspace*{-1\baselineskip}

\end{document}